\documentclass[prd,preprint,superscriptaddress,showpacs,nofootinbib,%
tightenlines ]{revtex4}
\usepackage{epsfig}
\usepackage{amssymb}
\newcommand{\gA}{\stackrel{\circ}{g}_{\!A}}

\newcommand{\ben}{\begin{displaymath}}
\newcommand{\een}{\end{displaymath}}
\newcommand{\be}{\begin{equation}}
\newcommand{\ee}{\end{equation}}
\newcommand{\bea}{\begin{eqnarray}}
\newcommand{\eea}{\end{eqnarray}}
\begin{document}
\title{
Improving the ultraviolet behavior in baryon chiral perturbation
theory}
\author{D.~Djukanovic}
\affiliation{Institut f\"ur Kernphysik, Johannes
Gutenberg-Universit\"at, D-55099 Mainz, Germany}
\author{M.~R.~Schindler}
\affiliation{Institut f\"ur Kernphysik, Johannes
Gutenberg-Universit\"at, D-55099 Mainz, Germany}
\author{J.~Gegelia}
\affiliation{Institut f\"ur Kernphysik, Johannes
Gutenberg-Universit\"at, D-55099 Mainz, Germany} \affiliation{High
Energy Physics Institute, Tbilisi State University, University
St.~9, 380086 Tbilisi, Georgia}
\author{S.~Scherer}
\affiliation{Institut f\"ur Kernphysik, Johannes
Gutenberg-Universit\"at, D-55099 Mainz, Germany}
\date{July 15, 2004}
\begin{abstract}
   We introduce a new formulation of baryon
chiral perturbation theory which improves the ultraviolet behavior
of propagators and can be interpreted as a smooth cutoff
regularization scheme. It is equivalent to the standard approach,
preserves all symmetries and therefore satisfies the Ward
identities. Our formulation is equally well defined in the vacuum,
one- and few-nucleon sectors of the theory. The equations
(Bethe-Salpeter, Lippmann-Schwinger, etc.) for the scattering
amplitudes of the few-nucleon sector are free of divergences in
the new approach. Unlike the usual cutoff regularization, our
'cutoffs' are parameters of the Lagrangian and do not have to be
removed.

\end{abstract}
\pacs{11.10.Gh,12.39.Fe}

\maketitle

\section{\label{introduction}Introduction}
   Weinberg's work in 1979 \cite{Weinberg:1979kz} originated
effective field theory (EFT) as one of the most important
theoretical tools for investigating strong-interaction processes
in the low-energy regime.
   The key progress due to Weinberg's approach was the development of a
perturbative scheme not in terms of a coupling constant, but
rather in terms of external momenta and the pion mass
\cite{Weinberg:1979kz}.
   In the traditional sense effective field theories are
non-renormalizable theories. However, as long as one includes all
of the infinite number of interactions allowed by symmetries, from
the point of view of removing divergences there is no difference
between the so-called non-renormalizable theories and
renormalizable theories \cite{Weinberg:mt}. Infinities encountered
in the calculation of loop diagrams are removed by a
renormalization of fields and the infinite number of free
parameters of the most general effective Lagrangian.

   The ideas of Weinberg were further developed and comprehensively
applied to the vacuum sector of QCD by Gasser and Leutwyler in
Refs.~\cite{Gasser:1984yg,Gasser:1984gg}.
   Chiral perturbation theory (ChPT) in the mesonic sector has been
successfully applied to calculations of various physical
quantities (for a recent review see, e.g., Ref.\
\cite{Scherer:2002tk}).
   Processes involving one nucleon in the initial and final states
were first considered by Gasser, Sainio, and \v{S}varc
\cite{Gasser:1988rb}.
   They observed that higher-loop diagrams can contribute to terms as
low as ${\cal O}(q^2)$, where $q$ generically denotes a small
expansion parameter such as, e.g., the pion mass.
   This problem has widely been interpreted as the absence of a systematic
power counting in the manifestly Lorentz-invariant formulation of
baryon chiral perturbation theory (BChPT).
   As an alternative the heavy-baryon formulation (HBChPT)
was suggested \cite{Jenkins:1990jv,Bernard:1992qa}.
   Most of the calculations in the one-baryon sector have been performed
in this framework using dimensional regularization in combination
with the modified minimal subtraction scheme ($\widetilde{\rm
MS}$) of ChPT (for an overview see, e.g., Refs.\
\cite{Scherer:2002tk,Bernard:1995dp}).
   The advantage of this approach is that it leads to a straightforward
power counting.
    Meanwhile it has been realized that, choosing an
appropriate renormalization condition, one can restore the power
counting in the original manifestly Lorentz-invariant formulation
of BChPT
\cite{Tang:1996ca,Ellis:1997kc,Becher:1999he,Gegelia:1999gf,
Gegelia:1999qt,Lutz:2001yb,Fuchs:2003qc,Fuchs:2003sh,Schindler:2003xv,Schindler:2003je}.

    A generalization to the few-nucleon sector was suggested in
Weinberg's papers on constructing nuclear forces from effective
field theory \cite{Weinberg:rz,Weinberg:um}.
   For processes involving $N>1$ nucleons, Weinberg proposed
applying the power counting to the potential, which is defined as
the sum of all $N$-nucleon-irreducible diagrams.
  The scattering amplitudes are then calculated by solving the
Lippmann-Schwinger (LS) or Schr\"odinger equation.

   The application of these ideas has encountered
various problems.
   They originate from the renormalization of the LS equation with
non-renormalizable potentials  (i.e. the iteration of the
potential generates divergent terms with structures which are not
included in the original potential). A consistent subtractive
renormalization requires the inclusion of the contributions of an
infinite number of counterterms which, in most cases, turns out to
be technically unfeasible. As a practical solution of the problem
one can perform the calculations in cutoff EFT. This approach
reproduces the results of the subtractively renormalized theory to
a given order, provided that the value of the cutoff parameter is
suitably chosen \cite{Lepage:1997cs,Gegelia:gn,Gegelia:1998iu,
Park:1998cu,Lepage:1999kt,Gegelia:2001ev,Gegelia:2004pz,Epelbaum:2004fk}.
    While this approach has been successful in various
applications
\cite{Ordonez:1995rz,Park:1997kp,Epelbaum:1999dj,Entem:2003ft,
Epelbaum:2004fk}, the applied cutoff regularization scheme breaks
certain symmetries of the theory and therefore special care has to
be taken. The application of cutoff regularization schemes to
effective theories has been of interest for a long time
\cite{Gasser:1979hf,Gasser:1980sb,Donoghue:1998rp,
Donoghue:1998bs,Espriu:1993if,Bernard:2003rp,Young:2002ib,
Leinweber:2003dg,Thomas:2003te}. A symmetry preserving lattice
regularization of ChPT in the presence of at most a single baryon
has been considered in Ref.~\cite{Borasoy:2003pg}. Although this
regularization could, in principle, also be applied in the
few-nucleon sector, to the best of our knowledge the question of
preserving symmetries in calculations of few-nucleon processes
still remains open.\footnote{So far, dimensional regularization
has only been used for a very restricted number of cases when the
equations are exactly solvable.}
 Therefore, the construction of a symmetry-preserving formulation
 of BChPT which renders equations free of divergences is of great interest.

   In this work we use an old idea by Slavnov \cite{Slavnov:aw} who
introduced chirally invariant terms with higher derivatives as a
regulator of the non-linear sigma model. We include
symmetry-preserving higher-derivative terms in the effective
Lagrangian of baryon chiral perturbation theory which modify the
ultraviolet behavior of the pion and baryon propagators. To
regularize the still remaining infinite number of primitively
divergent diagrams \cite{Slavnov:aw} we apply dimensional
regularization. This ensures that {\it all} loop diagrams are
regulated. The advantage of this approach is that it can be
applied to individual Feynman diagrams as well as to equations of
the few-nucleon sector.

  Our work is organized as follows.
  In Sec.\ \ref{effective_Lagrangian} we provide the terms which we add to the
standard effective Lagrangian.
     The nucleon mass is calculated within our new approach
in Sec.\ \ref{nucleon_self_energy}.
    In Sec.\ \ref{wardidentity} we demonstrate that the new scheme satisfies the U(1) Ward identity, while
    in Sec.\ \ref{nshbchpt} it is shown that in HBChPT, analogously
to the manifestly Lorentz-invariant formulation, the existence of
a consistent power counting depends on the applied renormalization
condition.
    Sec.\ \ref{NN} considers an application to simple examples of
the nucleon-nucleon scattering problem.
    A summary is given in Sec.\ \ref{conclusions},
while the Appendix contains the expressions for the required loop
integrals.

\section{The modified effective Lagrangian}
\label{effective_Lagrangian}

   The standard effective Lagrangian consists of the sum of the purely mesonic
and the $\pi N$, $NN$,  etc. Lagrangians, respectively,
\begin{equation}
{\cal L}_{\rm eff}={\cal L}_{\pi}+{\cal L}_{\pi N}+{\cal
L}_{NN}+\cdots. \label{inlagr}
\end{equation}
The terms in Eq.~(\ref{inlagr}) are organized in a (chiral)
derivative and quark-mass expansion
\cite{Weinberg:1979kz,Gasser:1984yg,Gasser:1984gg,Gasser:1988rb,%
Fearing:1994ga,Bijnens:1999sh,Fettes:2000gb,%
Ebertshauser:2001nj,Bijnens:2001bb}.
   Counting the quark-mass term as ${\cal O}(q^2)$
\cite{Gasser:1984yg,Colangelo:2001sp}, the mesonic Lagrangian
contains only even powers, whereas the baryonic Lagrangian
involves both even and odd powers due to the additional spin
degree of freedom. We choose to not show the counterterms
explicitly. Instead we accompany the Feynman rules with the
subtraction rules within a fixed renormalization condition. In
particular, we use the extended on-mass-shell (EOMS)
renormalization of Ref.~\cite{Fuchs:2003qc}.

    The lowest-order mesonic Lagrangian reads \cite{Gasser:1984yg}
\begin{equation}
\label{l2} {\cal L}_2=\frac{F^2}{4}\mbox{Tr}\left[D_\mu U \left(
D^\mu U\right)^\dagger\right] +\frac{F^2}{4}\mbox{Tr} \left( \chi
U^\dagger+ U\chi^\dagger \right),
\end{equation}
where $U$ is a unimodular unitary $(2\times 2)$ matrix containing
the Goldstone boson fields. The covariant derivative is defined as
$$
D_\mu U=\partial_\mu U-i r_\mu U+i U l_\mu,
$$
where
$$r_\mu=v_\mu+a_\mu, \ \ \ l_\mu=v_\mu-a_\mu, \ \ \ \chi =2 B (s+i p). $$
Here, $v_\mu$, $a_\mu$, $s$, and $p$ are external vector,
axial-vector, scalar, and pseudo-scalar sources, respectively.
   In Eq.\ (\ref{l2}), $F$ denotes the pion-decay constant in the chiral
limit: $F_\pi=F[1+{\cal O}(\hat{m})]=92.4$ MeV.
   We work in the isospin-symmetric limit $m_u=m_d=\hat{m}$,
and the lowest-order expression for the squared pion mass is
$M^2=2 B \hat{m}$, where $B$ is related to the quark condensate
$\langle \bar{q} q\rangle_0$ in the chiral limit
\cite{Gasser:1984yg}.

  In the nucleon sector, let
\begin{displaymath}
\Psi=\left(\begin{array}{c}p\\n\end{array}\right)
\end{displaymath}
denote the nucleon field with two four-component Dirac fields, $p$
and $n$, describing the proton and neutron, respectively.
   The most general $\pi N$ Lagrangian is bilinear in
$\bar\Psi(x)$ and $\Psi(x)$ and involves the quantities $u$,
$u_\mu$, $\Gamma_\mu$, $v_\mu^{(s)}$ and $\chi_{\pm}$ (and their
derivatives), which are defined as
$$
u^2=U,\quad u_\mu =iu^{\dagger} D_\mu U u^{\dagger},\quad
\Gamma_{\mu}=\frac{1}{2}\left[u^\dagger\partial_{\mu}u
+u\partial_{\mu}u^\dagger- i(u^{\dagger}r_{\mu}u+u
l_{\mu}u^{\dagger})\right],
$$
$$
\chi_{\pm}=u^{\dagger}\chi u^\dagger\pm u\chi^\dagger u.
$$
   In terms of these building blocks the lowest-order Lagrangian reads
\cite{Scherer:2002tk,Gasser:1988rb}
\begin{equation}
{\cal L}_{\pi N}^{(1)}=\bar \Psi \left( i\gamma_\mu D^\mu -m
+\frac{1}{2} \gA\gamma_\mu \gamma_5 u^\mu\right) \Psi,
\label{lolagr}
\end{equation}
   where $D_\mu\Psi = (\partial_\mu +\Gamma_\mu-i v^{(s)}_\mu)\Psi $
denotes the covariant derivative. (In the definition of the
covariant derivative we follow Ref.\ \cite{Ecker:1995rk}, where
$\Gamma_\mu$ only contains traceless external fields and the
coupling to the isosinglet vector field $v_\mu^{(s)}$ is
considered separately.) In Eq.~(\ref{lolagr}), $m$ and $\gA$ refer
to the chiral limit of the physical nucleon mass and the
axial-vector coupling constant, respectively.

Below we will calculate the nucleon self-energy to third order.
For that purpose, we will need one of the seven structures of the
Lagrangian at ${\cal O}(q^2)$ \cite{Gasser:1988rb,Fettes:2000gb},
\begin{equation}
{\cal L}_{\pi N}^{(2)} = c_1 \mbox{Tr}(\chi_{+})\bar\Psi\Psi
+\cdots. \label{p2olagr}
\end{equation}
The Lagrangian ${\cal L}^{(3)}_{\pi N}$ does not contribute to the
nucleon mass at the given order.

\medskip
\medskip

    To improve the ultraviolet behavior of the propagators
generated by the Lagrangian of Eq.~(\ref{inlagr}) we introduce
additional terms into the Lagrangian which modify the propagators
of the pion and the nucleon. In particular, we consider the
modified pion propagator
\begin{equation}
\Delta^\Lambda_\pi(p)= \frac{1}{p^2-M^2+i 0^+} \
\prod\limits_{j=1}^{N_\pi} \frac{\Lambda_{\pi j}^2}{\Lambda_{\pi
j}^2-p^2-i 0^+} \label{genregpprop}
\end{equation}
and the modified nucleon propagator
\begin{equation}
S_N^\Lambda (p)=\frac{1}{\left(
p\hspace{-.45em}/\hspace{.1em}-m+i0^+\right)} \
\prod\limits_{j=1}^{N_\Psi} \frac{\Lambda_{\Psi j
}^2}{\Lambda_{\Psi j}^2+m^2-p^2-i0^+}. \label{genregfprop}
\end{equation}
Here, $\Lambda_{\pi i}$ and $\Lambda_{\Psi j}$ are (independent)
parameters.\footnote{In the following we let $\Lambda$
collectively represent the $\Lambda_{\pi i}$ and $\Lambda_{\Psi
j}$.} For simplicity we use the standard prescription for dealing
with poles in the $\Lambda$-dependent factors of the modified
propagators. For sufficiently large values of the parameters
$\Lambda$ any other prescription leads to the same results for
low-energy physical quantities.
    The propagators above can be generated by a Lagrangian which, in
addition to the standard BChPT Lagrangian of Eq.~(\ref{inlagr}),
contains additional {\it symmetry-preserving} terms. These terms
vanish in the limit $\Lambda_{\pi i}\to \infty$, $\Lambda_{\Psi
j}\to \infty$.

    The choice of the additional terms of the Lagrangian is not
unique. Furthermore these terms not only generate the above
propagators, but also result in additional interaction terms. Our
choice is motivated by the simplicity of calculations. For the
pion sector we choose
$$
{\cal L}_{\pi\pi}^{\rm reg}= \sum\limits_{n=1}^{N_\pi}
\frac{X_n}{4} \ \frac{F_0^2}{4} \ {\rm Tr} \left( \left\{
\left(D^2\right)^nU U^\dagger -U\left[
\left(D^2\right)^nU\right]^\dagger \right\} \left[  D^2U
U^\dagger-U\left(D^2 U\right)^\dagger -\chi U^\dagger
+U\chi^\dagger \right] \right),
$$
where $D^2U=D_\alpha D^\alpha U$ and $X_n$ are functions of
$\Lambda_{\pi i}$. For example, in order to generate the modified
propagator
\begin{equation}
\Delta^\Lambda_\pi(p)= \frac{1}{p^2-M^2+i 0^+} \
\prod\limits_{j=1}^{3} \frac{\Lambda_{\pi j}^2}{\Lambda_{\pi
j}^2-p^2-i 0^+}, \label{regpprop}
\end{equation}
we need to take $N_\pi =3$ and
\begin{eqnarray}
X_1 & = & \frac{1}{\Lambda_{\pi 1}^2} +\frac{1}{\Lambda_{\pi
2}^2}+\frac{1}{\Lambda_{\pi 3}^2},
\nonumber \\
X_2 & = &  \frac{\Lambda^2_{\pi 1}+\Lambda^2_{\pi 2}+
\Lambda^2_{\pi 3}}{\Lambda^2_{\pi 1} \Lambda^2_{\pi 2}
\Lambda^2_{\pi 3}},    \nonumber \\
X_3 & = & \frac{1}{\Lambda^2_{\pi 1} \Lambda^2_{\pi
2}\Lambda^2_{\pi 3}}. \label{xis}
\end{eqnarray}

    For the additional terms of the Lagrangian of the nucleon sector we choose
\begin{equation}
{\cal L}_{\pi N}^{\rm reg}=\sum\limits_{n=1}^{N_\Psi}
\frac{Y_n}{2} \ \left[ \bar \Psi \left( i\gamma_\mu D^\mu -m
\right) \left( D^2+m^2\right)^n \Psi+h.c.\right],\label{Nlagrreg}
\end{equation}
where $Y_n$ are functions of $\Lambda_{\Psi j}$. For example, for
the modified nucleon propagator
\begin{equation}
S_N^\Lambda (p)=\frac{\Lambda_\Psi^2}{\left(
p\hspace{-.45em}/\hspace{.1em}-m+i 0^+\right)\left(
\Lambda_\Psi^2+m^2-p^2-i 0^+\right)} \label{regfprop}
\end{equation}
we have $N_\Psi=1$ and $Y_1=1/\Lambda^2_\Psi$.

    Depending on the order of the performed calculations
we choose the modified propagators, i.e.~fix $N_\pi$ and $N_\Psi$,
such that all loop diagrams (except some of the primitively
divergent diagrams) contributing to the given order converge. To
obtain the modified Lagrangian for lower order calculations one
needs to take $\Lambda_{\pi i}\to \infty$, $\Lambda_{\Psi j}\to
\infty$ for some of the parameters $\Lambda$ in the modified
Lagrangian used in higher-order calculations.

    Analogously to the nonlinear sigma model \cite{Slavnov:aw} the
additional terms do not render all loop diagrams finite. There
still remain an infinite number of primitively divergent diagrams
in the mesonic sector as well as divergences in diagrams with
fermion loops. These diagrams can be regularized in a
symmetry-preserving way by introducing additional auxiliary fields
analogously to the case of Yang-Mills theory \cite{Faddeev:be}.
However, in practical calculations such a technique is rather
difficult to apply. Instead it is possible (and much more
convenient) to use standard dimensional regularization. This is
due to the fact that the remaining divergent diagrams contribute
either in physical quantities of the vacuum (purely mesonic) and
the one-nucleon sectors, or they appear as sub-diagrams in the
potentials of the few-nucleon sector. In both cases the
calculations are perturbative, i.e., to any given order in the
chiral expansion one needs to calculate a finite number of
diagrams.
    Therefore, divergences which show up as the $1/(n-4)$
poles (where $n$ denotes the number of space-time dimensions) can
be explicitly subtracted (i.e.~absorbed in the redefinition of the
parameters of the effective Lagrangian).

    To summarize, our scheme consists of adding
symmetry-preserving additional terms in the standard effective
Lagrangian and applying dimensional regularization to the
resulting effective theory. All symmetries are preserved in the
regularized theory, i.e., regularized quantities satisfy all
relevant Ward identities. We expand the regularized diagrams in
powers of $n-4$ and subtract $1/(n-4)$ pole-terms observing that
there is a finite number of them to any given (finite) order in
the chiral expansion of physical quantities in the vacuum and one
nucleon sectors, and the potentials in the few-nucleon sector. No
further divergences occur (for finite parameters $\Lambda$)
neither in the vacuum and one-nucleon sector nor in the equations
of the few-nucleon sector. Therefore, for the equations of the
few-nucleon sector we can take $n=4$.

Using a field transformation the additional higher-derivative
terms which we introduced in the effective Lagrangian can be
reexpressed in a canonical form, i.e. a form with a minimal number
of independent terms \cite{Scherer:1994wi,Fearing:1999fw}. This
clearly shows that any $\Lambda$ dependence of the physical
quantities can systematically be absorbed in the redefinition of
the parameters of the standard canonical effective Lagrangian.

\section{\label{nucleon_self_energy}Nucleon self-energy}

As an example of the application of our approach we calculate the
nucleon self-energy to order ${\cal O}\left(q^3\right)$ in this
section. For this calculation it is sufficient to take
$N_\pi=N_\Psi=1$. We parametrize the complete nucleon propagator
as
\begin{equation}
S_N(p)= \frac{1}{\left( p\hspace{-.45em}/\hspace{.1em}-m+i
0^+\right)\left[ 1- \left( p^2-m^2\right)/\Lambda_\Psi^2\right]
-\Sigma(p\hspace{-.45em}/\hspace{.1em})}, \label{prpar}
\end{equation}
where $m$ is the nucleon pole mass in the chiral limit and the
nucleon self-energy $-i\Sigma(p\hspace{-.45em}/\hspace{.1em})$
represents the sum of all one-particle-irreducible perturbative
contributions to the two-point function. The physical nucleon mass
is defined through the pole of the full propagator at
$p\hspace{-.45em}/\hspace{.1em}=m_N$,
\begin{equation}
\label{nucleonmassdefinition} \left(
m_N-m\right)\left(1-\frac{m_N^2-m^2}{\Lambda_\Psi^2}
\right)-\Sigma(m_N)=0.
\end{equation}

\begin{figure}
\epsfig{file=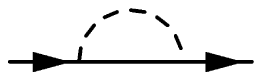,width=0.3\textwidth}
\caption[]{\label{nse:fig} One-loop contribution to the nucleon
self-energy at ${\cal O}(q^3)$.}
\end{figure}

\noindent
    At ${\cal O}(q^3)$, the self-energy receives contact
contributions from ${\cal L}_{\pi N}^{(2)}$ as well as the
one-loop contribution of Fig.\ \ref{nse:fig},
\begin{equation}
\Sigma=\Sigma_{\rm contact}+\Sigma_{\rm loop},
\end{equation}
where $\Sigma_{\rm contact}=-4 c_1 M^2$. Applying Feynman rules we
obtain for the one-loop contribution
\begin{eqnarray}
\label{sigmaaresult} \Sigma_{\rm loop} &=&-\frac{3 \gA^2
\Lambda_\pi^2\Lambda_\Psi^2}{4 F^2}\left\{
(p\hspace{-.45em}/\hspace{.1em}+m) \
I(1011) +M^2(p\hspace{-.45em}/\hspace{.1em}+m) \  I(1111) \right.\nonumber\\
&&\left. +(p^2-m^2)p\hspace{-.45em}/\hspace{.1em} \ I^{(p)}(1111)-
p\hspace{-.45em}/\hspace{.1em} \ I^{(p)}(1110)\right\},
\end{eqnarray}
where
\begin{equation}
\left\{ I(abcd), p^\mu I^{(p)}(abcd) \right\}=i \int \frac{d^4
k}{(2 \pi)^4} \frac{\left\{1,k^\mu\right\}}{A^a B^b C^c D^d},
\label{intdefIabcd}
\end{equation}
with
\begin{eqnarray}
&& A = k^2-\Lambda_\pi^2+i0^+, \nonumber \\ && B = k^2-M^2+i0^+,
\nonumber \\ && C = (p+k)^2-m^2-\Lambda_\Psi^2+i0^+, \nonumber \\
&& D = (p+k)^2-m^2+i0^+.\nonumber
\end{eqnarray}

To further simplify the calculation we take
$\Lambda_\pi=\Lambda_\Psi$. The parameter $n$ of dimensional
regularization has been put  $n=4$, as the diagram is finite for
finite $\Lambda$. We perform the renormalization by applying the
extended on mass-shell (EOMS) scheme of Ref.~\cite{Fuchs:2003qc}.
First we substitute the expressions for the loop integrals from
the Appendix and expand Eq.~(\ref{sigmaaresult}) in a power series
in $\Lambda$ (around $\Lambda =\infty$). We then subtract all
positive powers of $\Lambda$ and $\ln(\Lambda/m)$.\footnote{Note
that, since our scheme respects all symmetries of the theory, the
Ward identities are satisfied separately in each order of the
expansion in powers of $\Lambda$.} Next we expand the remaining
expression in powers of small quantities, i.e., $M$, $p^2-m^2$ and
$p\hspace{-.45em}/\hspace{.1em}-m$ and subtract all terms of
zeroth, first and second order in this expansion, so that the
renormalized expression is indeed of order $q^3$ as mandated by
the power counting. The resulting expression for the subtraction
terms reads
$$
\Sigma^{\rm sub}=-\frac{3 \gA^2\Lambda^2 \left( 4 m + 5
p\hspace{-.45em}/\hspace{.1em}\right)}{256 \pi^2 F^2}
+\frac{\gA^2}{256 \pi^2 m F^2} \ \Biggl[ -12 m^4 + 3 m^2 M^2 + 8
m^2 p^2 - 6 (p^2)^2
$$$$
- 10 m^3 p\hspace{-.45em}/\hspace{.1em}+ m p^2
p\hspace{-.45em}/\hspace{.1em} + 12 m \left( 2 m^3 + 4 m M^2 + 3
m^2 p\hspace{-.45em}/\hspace{.1em} - p^2
p\hspace{-.45em}/\hspace{.1em}\right)\ln\left(\frac{\Lambda}{m}\right)\Biggr]
$$
$$
-\frac{\gA^2\ m}{2560 \pi^2 F^2 \Lambda^2} \Biggl[ -48 m^4 - 10
(p^2)^2 + 6 m^2 \left( 15 M^2 + 41 p^2\right) + 63 m^3
p\hspace{-.45em}/\hspace{.1em} + 130 m p^2
p\hspace{-.45em}/\hspace{.1em}
$$\begin{equation}
- 240 m^2 \left( m^2 + 2 M^2 + p^2 + 2
mp\hspace{-.45em}/\hspace{.1em}\right) \ln
\left(\frac{\Lambda}{m}\right)\Biggr]. \label{st}
\end{equation}
Subtracting Eq.~(\ref{st}) from Eq.~(\ref{sigmaaresult}) and
taking $p\hspace{-.45em}/\hspace{.1em}=m_N$, we obtain for the
renormalized on-mass-shell self-energy to order $q^3$
$$
\Sigma^R|_{p\hspace{-.45em}/\hspace{.1em}=m_N} =-\frac{3 \gA^2
M^3}{32 \pi F^2}+{\cal O}\left(\frac{1}{\Lambda^4}\right).
$$
Using Eq.~(\ref{nucleonmassdefinition}), the nucleon mass to order
$q^3$ follows as
\begin{equation}
m_N=m - 4 c_1  M^2 -\frac{3\ \gA^2}{32 \pi F^2} \ M^3+{\cal
O}\left( \frac{1}{\Lambda^4}\right), \label{regmass}
\end{equation} which agrees with the standard BChPT
result
\cite{Becher:1999he,Fuchs:2003qc,Steininger:1998ya,Kambor:1998pi}.

\section{\label{wardidentity} Electromagnetic Ward identity}

To demonstrate that the new formulation indeed respects the
symmetries of the theory, we analyze the electromagnetic Ward
identity for the nucleon which, in units of the elementary charge,
reads
\begin{equation}
\left( p_f-p_i\right)_\mu \Gamma_{N}^{\mu} \left( p_f,
p_i\right)=\frac{1+\tau_3}{2} \ \left[ S^{-1}_N\left( p_f\right)-
S^{-1}_N\left( p_i\right) \right]. \label{ward}
\end{equation}
Here,
\begin{equation}
\Gamma^\mu_N \left( p_f, p_i\right)=\Gamma^\mu_{N 0}\left( p_f,
p_i\right) +\Lambda^\mu_N\left( p_f, p_i\right)\label{vertex}
\end{equation}
is the one-particle-irreducible three-point function ($\Psi {\cal
J}^\mu \bar\Psi$) with ${\cal J}^\mu$ the electromagnetic current
operator in units of the elementary charge. $\Gamma^\mu_{N0}\left(
p_f, p_i\right)$ corresponds to the tree-order contribution and
$\Lambda^\mu_N\left( p_f, p_i\right)$ consists of loop
corrections. In order to determine  $\Gamma_N^\mu$, we consider
the coupling to an external electromagnetic field ${\cal A_\mu}$
and insert for the external fields in Eq.~(\ref{lolagr})
\begin{displaymath}
r_\mu=l_\mu=-e \frac{\tau_3}{2}{\cal A_\mu},\quad
v_\mu^{(s)}=-\frac{e}{2} {\cal A_\mu}.
\end{displaymath}
   For the purpose of this section it is sufficient to take $N_\pi=N_\Psi=1$.
   From our modified Lagrangian we obtain
\begin{equation}
\Gamma^\mu_{N 0} \left( p_f, p_i\right) =\frac{1+\tau_3}{2} \
\gamma^\mu-\frac{1+\tau_3}{2} \ \frac{1}{2 \Lambda^2}\left[
\gamma^\mu \left(p_f^2+p_i^2-2 m^2\right)+\left(
p_f+p_i\right)^\mu \left(
p\hspace{-.45em}/\hspace{.1em}_f+p\hspace{-.45em}/\hspace{.1em}_i-2
m\right) \right], \label{frapbp}
\end{equation}
where $\Gamma^\mu_{N0} \left( p_f, p_i\right)$ 
and the free propagator of Eq.~(\ref{regfprop}) satisfy the
relation
\begin{equation}
\left( p_f-p_i\right)_\mu \Gamma^\mu_{N0} \left( p_f,
p_i\right)=\frac{1+\tau_3}{2}\left[ {S_N^\Lambda}^{-1}\left(
p_f\right)- {S_N^\Lambda}^{-1}\left(
p_i\right)\right].\label{wardtree}
\end{equation}
Of course, this result is not surprising, because the coupling to
an external electromagnetic field in the Lagrangian of Eq.\
(\ref{Nlagrreg}) proceeds via covariant derivatives which
essentially amount to a minimal coupling. At tree level this
automatically results in contributions satisfying the Ward
identity.\footnote{In the context of EFT the use of
minimal-substitution terms alone is not sufficient to generate a
consistent framework, because the {\em most general} effective
Lagrangian also contains terms involving field-strength tensors
such as, e.g., the $l_5$ and $l_6$ terms of ${\cal L}_4$
\cite{Gasser:1984yg}. In general, the presence of these terms is
also necessary for the purposes of renormalization (see Ref.\
\cite{Koch:2001ii} for a critical discussion of this issue).}

\begin{figure}
\epsfig{file=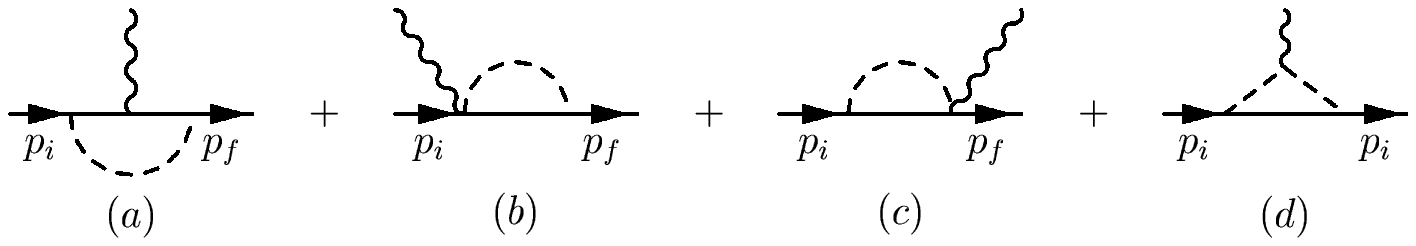, width=\textwidth}
\caption[]{\label{mnloops:fig} One-loop contributions to the
electromagnetic vertex.}
\end{figure}

    For the one-loop corrections to the nucleon self-energy
(of Fig.~\ref{nse:fig}) and the vertex (diagrams of
Fig.~\ref{mnloops:fig}) we obtain
\begin{eqnarray}
\label{sigmaa} \Sigma_{\rm loop}
(p\hspace{-.45em}/\hspace{.1em})&=&\frac{3 \gA^2}{4
F^2}\,i\int\frac{d^n k}{(2\pi)^n}\,
k\hspace{-.45em}/\hspace{.1em}\gamma_5 \ S_N^{\Lambda}\left(
p+k\right) k\hspace{-.45em}/\hspace{.1em}\gamma_5\,
\Delta_{\pi}^\Lambda (k),
\\
\Lambda_{Na}^\mu \left( p_f, p_i\right)&=&\frac{\gA^2}{4
F^2}\,i\int\frac{d^n k}{(2\pi)^n}\,
k\hspace{-.45em}/\hspace{.1em}\gamma_5 \
\tau^a \ S_N^{\Lambda} \left( p_f+k\right)\nonumber \\
& & \times \Gamma^\mu_{N0}\left(p_f+k,p_i+k\right) S_N^{\Lambda}
\left( p_i+k\right) k\hspace{-.45em}/\hspace{.1em}\gamma_5\,
\tau^a \
\Delta_{\pi}^\Lambda (k), \label{vcora} \\
\Lambda_{Nb}^\mu \left( p_f, p_i\right)&=&2  \tau_3 \
\frac{\gA^2}{4 F^2}\,i\int\frac{d^n k}{(2\pi)^n}\,
k\hspace{-.45em}/\hspace{.1em}\gamma_5 \ S_N^{\Lambda} \left(
p_f+k\right) \gamma^\mu\gamma_5\,
\Delta_{\pi}^\Lambda (k), \label{vcorb} \\
\Lambda_{Nc}^\mu \left( p_f, p_i\right)&=& 2 \tau_3 \
\frac{\gA^2}{4 F^2}\,i\int\frac{d^n k}{(2\pi)^n}\, \gamma^\mu
\gamma_5 \ S_N^{\Lambda} \left( p_i+k\right)
k\hspace{-.45em}/\hspace{.1em}\gamma_5\,
\Delta_{\pi}^\Lambda (k), \label{vcorc} \\
\Lambda_{Nd}^\mu \left( p_f, p_i\right)&=& 2 \tau_3 \
\frac{\gA^2}{4 F^2}\,i\int\frac{d^n k}{(2\pi)^n}\,
({p\hspace{-.45em}/\hspace{.1em}}_i+k\hspace{-.45em}/\hspace{.1em}
-{p\hspace{-.45em}/\hspace{.1em}}_f)
\gamma_5 \ S_N^{\Lambda}\left(p_i+k\right)\nonumber\\
&&\times\Gamma^\mu_{\pi 0}\left(p_f-p_i-k,-k\right)
k\hspace{-.45em}/\hspace{.1em}\,\gamma_5\, \Delta_{\pi}^\Lambda
(k+p_i-p_f) \Delta_{\pi}^\Lambda (k), \label{vcord}
\end{eqnarray}
where
\begin{equation}
\Gamma^\mu_{\pi 0}(p',p)=\left(p'+p\right)^\mu
\left(1-\frac{p'^2+p^2-M^2}{\Lambda_\pi^2}\right) \label{pvfr}
\end{equation}
is the leading tree-order contribution in $\Gamma^\mu_\pi$, which
is related to the one-particle-irreducible three-point function
($\pi_j {\cal J}^\mu \pi_i$) by the relation
\begin{equation}
\Gamma_{\pi ji}^\mu(p',p)=i \epsilon_{3ij} \Gamma^\mu_{\pi
}(p',p). \label{vr}
\end{equation}
    To check the Ward identity for the above loop diagrams, we multiply
$\Lambda^\mu_N=\Lambda^\mu_{Na}+\Lambda^\mu_{Nb}+\Lambda^\mu_{Nc}+\Lambda^\mu_{Nd}$
with $\left(p_f-p_i\right)_\mu$, use Eq.~(\ref{wardtree}) and the
Ward identity for pions (at leading tree order)
\begin{equation}
\left( p'-p\right)_\mu \Gamma^\mu_{\pi
0}(p',p)={\Delta^\Lambda_\pi}^{-1} (p')-{\Delta^\Lambda_\pi}^{-1}
(p),\label{pwi}
\end{equation}
and obtain after a straightforward calculation
\begin{equation}
\left( p_f -p_i\right)_\mu \Lambda^\mu_N \left( p_f,
p_i\right)=\frac{1+\tau_3}{2} \ \left[\Sigma_{\rm
loop}({p\hspace{-.45em}/\hspace{.1em}}_i)- \Sigma_{\rm
loop}({p\hspace{-.45em}/\hspace{.1em}}_f)\right], \label{wi1l}
\end{equation}
which verifies the Ward identity of Eq.~(\ref{ward}).

\section{\label{nshbchpt} Nucleon self-energy diagram in HBChPT}

 It is common practice to assume the existence of
a consistent power counting in HBChPT without specifying the
renormalization scheme used. In HBChPT, as in any quantum field
theory, one has the freedom to choose a renormalization condition.
Dimensional regularization in combination with the $\widetilde{\rm
MS}$ scheme, which is commonly used in HBChPT, is only one among
an infinite number of possibilities.
    In this section we apply our higher-derivative formulation  to the
nucleon self-energy diagram of Fig.~\ref{nse:fig} in order to show
that, analogously to the manifestly Lorentz-invariant formulation,
the existence of a consistent power counting in HBChPT depends on
the choice of the renormalization condition.

    Using the pion propagator of Eq.~(\ref{genregpprop}) for $N_\pi=2$
and $\Lambda_{\pi 1}=\Lambda_{\pi 2}=\Lambda$, we obtain (see,
e.g., Section 5.5.9 and Appendix C.2 of Ref.\
\cite{Scherer:2002tk} for a detailed calculation in HBChPT)
\begin{equation}
\Sigma^{(3)}_{\rm loop}(p)=3 \ \frac{\gA^2\Lambda^4}{F^2} \
S_\mu^v S_\nu^v \ J^{\mu\nu}_{\pi N}(121;\omega),
\label{hbchptfse}
\end{equation}
where $\omega=(p\cdot v-m)$ and
\begin{equation}
J^{\mu\nu}_{\pi N}(abc;\omega)= i \int\frac{d^4 k}{(2
\pi)^4}\frac{k^\mu k^\nu }{\left[ k^2-M^2+ i
0^+\right]^a\left[k^2-\Lambda^2+ i 0^+\right]^b\left[ v\cdot
k+\omega+ i 0^+\right]^c}. \label{Jpnotdef}
\end{equation}
One can parameterize $J^{\mu\nu}_{\pi N}(121;\omega)$ as
\begin{equation}
J^{\mu\nu}_{\pi N}(121;\omega)= c_1 g^{\mu\nu}+c_2 v^\mu v^\nu.
\label{Jpnotdef11}
\end{equation}
Since $S_v\cdot v=0$, $c_2$ does not contribute to the
self-energy. For $c_1$ we find
\begin{equation}
c_1=\frac{1}{3} \left[ \left( M^2-\omega^2\right) J_{\pi N
}(121;\omega)+ J_{\pi N}(021;\omega)+\omega J_{\pi N}(120;\omega)
\right], \label{c1}
\end{equation}
where
\begin{equation}
J_{\pi N}(abc;\omega)= i \int \frac{d^4 k }{(2
\pi)^4}\frac{1}{\left[k^2-M^2+ i 0^+\right]^a\left[ k^2-\Lambda^2+
i 0^+\right]^b\left[ v\cdot k+\omega+ i 0^+\right]^c}.
\label{Jpnodef}
\end{equation}

    Standard power counting of HBChPT assigns the order ${\cal O}\left(q^3\right)$
to the diagram of Fig.~\ref{nse:fig}. Calculating the loop
integrals of Eq.~(\ref{Jpnodef}) (see Appendix \ref{hbchpt}), we
obtain
\begin{equation}
\Sigma^{(3)}_{\rm loop}(p)= -\frac{3 \gA^2}{64 \pi^2 F^2} \ \left[
\pi \Lambda^3 + 3 \ \omega \ \Lambda^2\right]+{\cal
O}(\Lambda).\label{hbchptfsecalc}
\end{equation}
Both terms inside the square brackets (as well as the term
proportional to $\Lambda$ which, for the sake of brevity, we have
not displayed) violate the power counting. They are analytic in
momenta and can be absorbed in the renormalization of the nucleon
mass and the nucleon field.
   Note that the
corresponding mass counterterm $\delta m \bar N_v N_v$, cancelling
the $\Lambda^3$ term in Eq.~(\ref{hbchptfsecalc}), is equal to
zero in the standard formulation of HBChPT with dimensional
regularization and is, therefore, usually not indicated in the
effective Lagrangian of HBChPT.\footnote{This is analogous to the
case of the pion tadpole self-energy in cutoff regularization,
where one needs a counterterm of order $p^2$ for the pion mass
\cite{Gasser:1979hf,Gasser:1997rx}.} Choosing the renormalization
scheme appropriately one can subtract {\it all} terms in
Eq.~(\ref{hbchptfsecalc}) which violate the power counting so that
the renormalized diagram is of order ${\cal O}\left( q^3\right).$

\section{\label{NN} NN sector}

\subsection{\label{contactinteraction}Contact interaction}
    In this section we consider a demonstrating example of the
application of our approach to the NN problem in the manifestly
Lorentz-invariant formulation of baryon chiral perturbation
theory.

    Let us consider the simplest possible
$( \bar\Psi\Psi)^2$ contact interaction term. The corresponding
equation for the lowest-order amplitude can be solved
analytically, therefore, one can apply standard dimensional
regularization to this problem. The interaction term in the
Lagrangian reads
\begin{equation}
{\cal L}_{NNNN}=C \ \bar\Psi\Psi \ \bar\Psi\Psi.
\label{NNLagrdreg}
\end{equation}

\begin{figure}
\epsfig{file=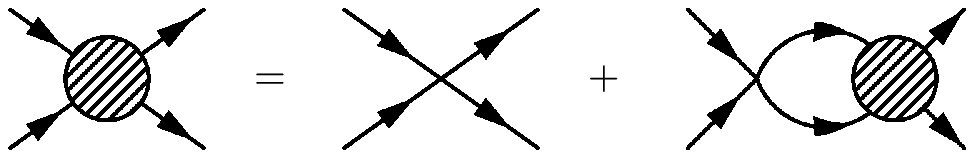,width=0.8\textwidth}
\caption[]{\label{nnampl:fig} Graphical illustration of the
equation for the NN scattering amplitude.}
\end{figure}

   In the following, we will consider the scattering amplitude in
the center-of-mass frame.
   Let $P=p_1+p_2$ denote the total four-momentum of the scattered
nucleons, where $p_1^\mu=(\sqrt{m^2+p^2},\vec p\,)$, $p_2^\mu
=(\sqrt{m^2+p^2},-\vec p\,)$ with $p=|\vec p\,|$ and $P^2=4 m^2+4
p^2$.

The interaction Lagrangian of Eq.~(\ref{NNLagrdreg}) generates the
following NN vertex (two-nucleon-irreducible contribution in the
NN scattering amplitude)
\begin{equation}
i V_{\lambda\sigma,\mu\nu}=2 i C \left(
\delta_{\lambda\nu}\delta_{\sigma\mu}
-\delta_{\lambda\mu}\delta_{\sigma\nu}\right). \label{NNpot}
\end{equation}

   According to Weinberg's approach, to find the corresponding
lowest-order NN scattering amplitude we need to solve the
equation\footnote {It is understood that
$T_{\lambda\sigma,\mu\nu}$ needs to be multiplied with the
corresponding Dirac spinors.}
\begin{equation}
T_{\lambda\sigma,\mu\nu}(P)=V_{\lambda\sigma,\mu\nu}+i \int
\frac{d^4 k}{(2 \pi)^4} \ V_{\lambda\sigma, \alpha\gamma} \
G_{\alpha\gamma,  \beta\delta} (P,k) \ T_{\beta\delta,\mu\nu}(P),
\label{Meq}
\end{equation}
schematically shown in Fig.~\ref{nnampl:fig}, where
\begin{eqnarray*}
G_{\alpha\gamma,\beta\delta}^{\rm dr}(P,k)&=&-\frac{\left( k_\mu
\gamma^\mu_{\alpha\beta}+m \delta_{\alpha\beta} \right) \left[
\left(P_\nu-k_\nu\right) \gamma^\nu_{\gamma\delta}+m
\delta_{\gamma\delta}\right] }{\left[ k^2-m^2+i 0^+\right]\left[
\left( P-k\right)^2-m^2+i 0^+\right]},\\
G_{\alpha\gamma,\beta\delta}^{\rm hd}(P,k)&=&\frac{\Lambda_\Psi^4}
{\left[ k^2-m^2-\Lambda_\Psi^2+i 0^+\right]\left[\left(
P-k\right)^2-m^2-\Lambda_\Psi^2+i 0^+\right]} \
G_{\alpha\gamma,\beta\delta}^{\rm dr}(P,k)
\end{eqnarray*}
are the (two-nucleon) propagators to be used in standard
dimensional regularization and higher-derivative formulation,
respectively.
   Integrating Eq.~(\ref{Meq}) over $k$, we obtain
\begin{equation}
T_{\lambda\sigma,\mu\nu}(P)=V_{\lambda\sigma,  \mu\nu}+ i
V_{\lambda\sigma, \alpha\gamma} \ {\cal G}_{\alpha\gamma,
\beta\delta} (P) \ T_{\beta\delta, \mu\nu}(P), \label{Meqmatrix}
\end{equation}
where \begin{eqnarray} {\cal G}_{\alpha\gamma, \beta\delta} (P)&=&
i \Biggl\{ m \delta_{\alpha\beta} \ \left(
\gamma^\mu_{\gamma\delta} P_\mu +m \delta_{\gamma\delta}\right)
I_{NN}\nonumber\\
&&+\left[ \gamma^\mu_{\alpha\beta} \left(
\gamma_{\gamma\delta}^\nu P_\nu +m \delta_{\gamma\delta}\right)-m
\delta_{\alpha\beta} \gamma^\mu_{\gamma\delta}\right]I^{(P)}_{NN}
P_\mu -\gamma^\mu_{\alpha\beta} \gamma^\nu_{\gamma\delta} I_{NN,
\mu\nu}\Biggr\}, \label{calg}
\end{eqnarray}
with
\begin{equation}
\left\{ I_{NN}, \ P^\mu I_{NN}^{(P)}, \
I_{NN}^{\mu\nu}\right\}^{\rm dr}=i \int \frac{d^n k }{( 2
\pi)^n}\,\frac{\left\{ 1, \ k^\mu, \ k^\mu k^\nu \right\}}{\left[
k^2-m^2+i 0^+\right]\left[ \left( P-k\right)^2-m^2+i 0^+\right]}
\label{intdefdr}
\end{equation}
in standard dimensional regularization and
\begin{eqnarray}
\left\{ I_{NN}, \ P^\mu I_{NN}^{(P)}, \
I_{NN}^{\mu\nu}\right\}^{\rm hd}&=&i \int \frac{d^4 k}{(2\pi)^4}
\frac{\left\{ 1, \ k^\mu, \ k^\mu k^\nu \right\}}{\left[ k^2-m^2+i
0^+\right]\left[ \left( P-k\right)^2-m^2+i 0^+\right]} \nonumber\\
&&\times \frac{\Lambda_\Psi^4}{\left[ k^2-m^2-\Lambda_\Psi^2+i
0^+\right]\left[ \left( P-k\right)^2-m^2-\Lambda_\Psi^2+i
0^+\right] } \label{intdef}
\end{eqnarray}
in higher-derivative formulation, respectively.

    We renormalize Eq.~(\ref{Meqmatrix}) by subtracting the
contributions of loop integrals at $P^2=4 m^2$. Next we expand the
subtracted loop integrals ($I^R=I-I|_{P^2=4 m^2}$) in $p$ and
retain terms to order ${\cal O}\left( p\right)$.  The resulting
equation reads
\begin{equation}
T^R_{\lambda\sigma, \mu\nu}(P)= V_{\lambda\sigma, \mu\nu}+i
V_{\lambda\sigma, \alpha\gamma} \ {\cal G}_{\alpha\gamma,
\beta\delta}^R \ T^R_{\beta\delta,  \mu\nu}(P),
\label{Meqmatrixren}
\end{equation}
where
$$
{\cal G}_{\alpha\gamma, \beta\delta}^R=\frac{p}{16 \pi m }\Biggl\{
m \delta_{\alpha\beta} \ \left( \gamma^\mu_{\gamma\delta} P_\mu +m
\delta_{\gamma\delta}\right)
$$
\begin{equation}
+ \frac{P_\mu}{2} \ \left[ \gamma^\mu_{\alpha\beta} \left(
\gamma_{\gamma\delta}^\nu P_\nu +m \delta_{\gamma\delta}\right)-m
\delta_{\alpha\beta}
\gamma^\mu_{\gamma\delta}\right]-\gamma^\mu_{\alpha\beta}
\gamma^\nu_{\gamma\delta} \ \frac{P^\mu P^\nu}{4} \Biggr\}.
\label{NNdiagram}
\end{equation}
   Comparing Eq.~(\ref{Meqmatrixren}) for standard dimensional regularization and
higher-derivative formulation (remembering that terms of order
${\cal O}(p^2)$ have been neglected) we obtain identical
results.\footnote{Although Eq.~(\ref{Meqmatrixren}) can be solved
exactly, it is beyond the scope of this paper to perform this
straightforward but rather cumbersome calculation.}

   Note that, in contrast to the example considered above,
it is not clear how to apply standard dimensional regularization
to equations involving potentials derived from BChPT. The
advantage of our higher-derivative formulation is that it is also
applicable in these cases while preserving the symmetries of the
theory.

\subsection{\label{onepionexchange}Inclusion of one-pion exchange potential}

Below we consider some conceptual issues of renormalization and
the $\Lambda$ dependence of the leading-order ${\rm NN}$
scattering amplitude in a non-relativistic formalism of BChPT.
This amplitude is obtained by solving the Lippmann-Schwinger
equation with a contact interaction plus the one-pion exchange
potential.\footnote{A detailed discussion of the heavy-baryon
reduction of our new Lagrangian (including a numerical analysis in
the few-body sector) will be given in a forthcoming publication.}

Using an appropriate field redefinition and the standard
heavy-baryon reduction with $v=(1,0,0,0)$ we obtain the
leading-order ${\rm NN}$ potential (for the choice $N_\Psi=1$,
$N_\pi=0$)
\begin{equation}
V\left(\vec p\,',\vec p\,\right)=C_S+C_T \ {\vec\sigma}_1\cdot
{\vec\sigma}_2-\left(\frac{\gA}{2 F}\right)^2 \ (\vec{\tau}_1
\cdot \vec{\tau}_2) \ \frac{{\vec\sigma}_1\cdot (\vec p\,'-\vec
p\,) \ {\vec\sigma}_2\cdot (\vec p\,'-\vec p\,)}{\left(\vec
p\,'-\vec p\,\right)^2+M_\pi^2}, \label{opepot}
\end{equation} \noindent
where $C_S$ and $C_T$  are the coupling constants of the
four-nucleon contact interaction Lagrangian at leading order. The
scattering amplitude satisfies the equation
\begin{equation}
T\left(\vec p\,',\vec p\,\right)=V\left(\vec p\,',\vec p\,\right)+
m \int \frac{d^3\vec k}{(2 \pi)^3} \ V\left(\vec p\,',\vec
k\right) \ \frac{\Lambda^2}{\left(\Lambda^2+\vec k^2
\right)\left(m E-\vec k^2+i 0^+\right)} \ T\left(\vec k,\vec
p\right), \label{opeampl}
\end{equation}
where $E={\vec p}\ ^2/m$ is the energy of two nucleons in the
center-of-mass system.

We fix the free parameters $C_S$ and $C_T$ as functions of
$\Lambda$ by demanding that the solution of equation
(\ref{opeampl}) reproduces two physical quantities for fixed
kinematics. Following Ref.~\cite{Weinberg:um} we choose the
renormalization points of the order of small external momenta.
This exactly corresponds to the following general renormalization
program (see, e.g., \cite{Gasser:1982ap}). First one calculates
the quantities of physical interest in terms of bare parameters in
the regularized theory. Once a sufficient number of physical
quantities are determined as functions of bare parameters, one
inverts the results and expresses the bare quantities in terms of
physical quantities. These expressions are then used to eliminate
the bare parameters in all other quantities of physical interest.
This procedure preserves all symmetries provided that the applied
regularization scheme respects them. If the considered theory is
renormalizable in the standard sense the above procedure removes
all divergences.

    For definiteness, let us take the zero-kinematics as
renormalization points.\footnote{This would not be a good choice
if we took the $\Lambda\to\infty$ limit in the end. We would be
faced with the problem of very poor convergence
\cite{Kaplan:1996xu}.} This corresponds to the subtraction of loop
diagrams at zero-kinematics. In the above case, expressing $C_S$
and $C_T$ from two physical quantities\footnote{We could take as
"quantities of physical interest" the scattering lengths of the ${
}^1S_0$ and ${ }^3S_1$ NN scattering.} and substituting them into
other quantities we only eliminate some of the terms that diverge
in the limit $\Lambda\to\infty$. This is due to the
non-renormalizability of BChPT in the traditional sense. However,
note that $\Lambda$ is a parameter of the Lagrangian and we do not
have to take it to infinity. The remaining $\Lambda$ dependence of
the amplitude is of higher order in the small-parameter expansion
(pion mass, small external momenta). As the potential of
Eq.~(\ref{opepot}) is non-renormalizable in the traditional sense,
the perturbative expansion of the renormalized amplitude contains
negative as well as positive powers of $\Lambda$ (and/or positive
powers of $\ln\Lambda$). These contributions contain terms of the
form
\begin{equation}
\sim \frac{q^i}{\Lambda^j} \ \ \ \ {\rm as \ well \ as} \ \ \sim
\frac{q^i\Lambda^j}{Q^{i+j}}, \ {\rm with} \ \ i>0, \ \ j>0,
\label{divterms}
\end{equation}
where $q$ denotes small external momenta or the pion mass and $Q$
stands for $4 \pi F$ and/or the large scale parameter hidden in
renormalized contact interaction constants. To keep these formally
higher-order contributions indeed suppressed numerically, one
should take $\Lambda\sim Q$. The existence of such an optimal
value of the parameter $\Lambda$ depends on the validity of the
assumption of Weinberg's approach that the renormalized
coupling constants are natural for renormalization points of the
order of or less than small external momenta. The validity of this
assumption has to be checked at each order of calculations. While
one cannot take the existence of the optimal value of $\Lambda$
for granted, the reasonable success of cutoff EFT suggests that it
should exist. A detailed analysis of this issue in our
symmetry-preserving approach is in progress.

The complete $\Lambda$ dependence of physical quantities can be
absorbed in the redefinition of couplings constants of the
canonical Lagrangian.\footnote{The original coupling constants
$c_i$ are written as $c_i=c_i^r+\delta c_i$, where the $c_i^r$ are
redefined coupling constants and the loop expansion of the $\delta
c_i$ part exactly cancels the corresponding $\Lambda$-dependent
parts of loop diagrams. The $c_i^r$ are independent of momenta,
i.e. local interaction terms of the effective Lagrangian remain
local.} For the above example this means that the contributions of
the form of Eq.~(\ref{divterms}) can be absorbed in the
redefinition of higher-order coupling constants. We could take any
value for the parameter $\Lambda$ provided that the compensating
contributions of higher-order terms (an infinite number of them)
are also included, but that does not seem to be feasible. Note
that $\Lambda$ is {\it not} a cutoff-regularization parameter and
does not need to be taken to infinity. The above specified optimal
choice of the {\it free parameter of the Lagrangian}, $\Lambda$,
ensures that, to the accuracy of the given calculations, physical
quantities do not depend on higher-order terms which we introduced
in the Lagrangian ($\Lambda$ independence).

As one cannot solve equations exactly, one carries out the above
renormalization  program numerically by fixing coupling constants
as functions of $\Lambda$ so that the given particular physical
quantities at the renormalization points (i.e. the fixed
$\Lambda$-independent values of them) are reproduced. The
reliability of this numerical renormalization procedure in
comparison with the explicit analytic renormalization depends only
on the accuracy of the numerical approximation, i.e. the two
approaches are conceptually equivalent.

In the approach suggested in this work, Ward identities are
satisfied order by order in the loop expansion as well as in the
chiral expansion of physical quantities \cite{Fuchs:2004nn}. In
the few-nucleon sector the physical quantities, at any finite
order in the chiral expansion, contain an infinite number of terms
in the loop expansion. On the other hand, to any specified order
$q^n$, for a processes involving $A$ nucleons, there is  a {\it
finite} number of $A$-nucleon irreducible diagrams. The sum of
these diagrams is defined as the effective potential. An infinite
number of diagrams contributing in the scattering amplitude (at
given order $q^n$) is summed up by solving the corresponding
equations with given $n$-th order effective potential. I.e.
substituting the $q^n$-order potential in the Lippmann-Schwinger
equation and performing the renormalization properly (as specified
above) corresponds to the summation of all {\it renormalized}
diagrams up to order $q^n$. The solution of the Lippmann-Schwinger
equation also contains some, but not all, of the higher-order
contributions, and the result is reliable only to order $q^n$, the
error being of order $O(q^{n+1})$. As the Ward identities are
satisfied order by order in the chiral expansion, and the
Lippmann-Schwinger equation resums {\it all} contributions to
order $q^n$, the contributions in physical quantities which can
violate the identities are of order $\sim O(q^{n+1})$, i.e. beyond
the accuracy of the given calculations.

\section{\label{conclusions}Summary}
    We have discussed a new formulation of BChPT,
which preserves all symmetries of the theory. The main idea is to
use some of the structures of the most general effective
Lagrangian to improve the ultraviolet behavior of propagators. The
coefficients of these terms depend on parameters (with dimension
of mass) which serve as smooth cutoffs of the theory. For
practical applications it is convenient to choose these parameters
to be equal.

    We have explicitly applied our new approach to a
calculation of the nucleon mass to order ${\cal O}\left(
q^3\right)$. We have also explicitly verified that the
electromagnetic Ward identities are satisfied by (strong)
one-loop-order corrections. The application of this scheme to the
one-nucleon sector of HBChPT demonstrates that the existence of a
consistent power counting scheme in HBChPT actually depends on the
applied renormalization scheme.

    The considerable advantage of the new formulation in
comparison with standard dimensional regularization is that, while
preserving all symmetries of the effective theory, it leads to
equations in the few nucleon (NN, NNN, etc.) sector which are free
of divergences. We have explicitly considered examples of the
contact interaction and one-pion exchange potentials in the NN
scattering problem and have discussed issues of renormalization
and consistency.


\acknowledgments

J.G. and M.R.S.~acknowledge the support of the Deutsche
Forschungsgemeinschaft (SFB 443).

\appendix
\section{\label{nse}Nucleon self-energy}

    The explicit expressions for the loop integrals of Eq.\
(\ref{intdefIabcd}) contributing in the calculation of the nucleon
self-energy up to and including order ${\cal
O}\left(\frac{1}{\Lambda^2}\right)$ read
\begin{eqnarray}
\Lambda^4 I(1111)&=&I_{\pi N}^f + \frac{1}{16 \pi^2} + \frac{1}{8
\pi^2} \
\ln\left(\frac{m}{\Lambda}\right) \nonumber\\
&& + \frac{1}{\Lambda^2} \ \left[ M^2  I_{\pi N}^f + \frac{5 p^2
-3 m^2+6 M^2}{96\ \pi^2}  \right. \nonumber\\
&&\texttt{} \left. + \frac{\left( m^2 + 2 M^2\right)
\ln\left(\frac{m}{\Lambda}\right)}{8\ \pi^2} + \frac{M^2\
\ln\left(\frac{M}{m}\right)}{8 \pi^2}\right], \label{J1111}
\end{eqnarray}
\begin{eqnarray}
\Lambda^4 I^{(p)}(1110) &=& -\frac{\Lambda^2}{64 \pi^2}
+\frac{m^2}{96
\pi^2}+ \frac{M^2}{64 \pi^2}- \frac{p^2}{192 \pi^2} \nonumber \\
&&+ \frac{1}{\Lambda^2} \ \Biggl[ \frac{3 M^4+M^2\ p^2}{64
\pi^2}-\frac{m^4}{128 \pi^2}-\frac{m^2 M^2}{48 \pi^2} +\frac{7 m^2
p^2-2 (p^2)^2}{960 \pi^2}\nonumber\\&&+ \frac{M^4}{16 \pi^2} \
\ln\left( \frac{M}{\Lambda}\right) \Biggr], \label{J1110p}
\end{eqnarray}
\begin{eqnarray}
\Lambda^4 I^{(p)}(1111) &=& \left( -\frac{1}{2} + \frac{m^2}{2
p^2} - \frac{ M^2}{2 p^2}\right) I^f_{\pi N} - \frac{1}{32 \pi^2}
- \frac{1}{16 \pi^2} \ \ln\left(\frac{m}{\Lambda}\right) +
\frac{M^2}{16
p^2\pi^2} \ \ln\left(\frac{M}{m}\right) \nonumber\\
&& + \frac{1}{\Lambda^2} \ \Biggl[ I^f_{\pi N} \left( -\frac
{M^2}{2} + \frac{ m^2 M^2}{2 p^2} - \frac{M^4}{2 p^2}\right) -
\frac{m^2}{48 \pi^2}\nonumber\\
&& - \frac{5 p^2}{192 \pi^2} - \frac{2 m^2+ M^2 }{16 \pi^2} \ \ln
\left(\frac{m}{\Lambda}\right) + \frac{M^4}{16 \pi^2 p^2} \
\ln\left(\frac{M}{m}\right)\Biggr], \label{J3111p}
\end{eqnarray}
\begin{eqnarray}
\Lambda^4 I(1011) &=& \frac{\Lambda^2}{16\ \pi^2} + \frac{m^2}{32\
\pi^2} + \frac{p^2}{48 \pi^2} + \frac{m^2}{8 \pi^2} \
\ln\left(\frac{ m}{\Lambda}\right) \nonumber\\
&& + \frac{1}{\Lambda^2} \ \Biggl[  -\frac{m^4}{96\ \pi^2} +
\frac{19\ m^2 p^2}{192 \pi^2} + \frac{3\ (p^2)^2}{320 \pi^2} +
\frac{m^2 \left(m^2+p^2\right)}{8 \pi^2} \
\ln\left(\frac{m}{\Lambda}\right) \Biggr], \label{J1011}
\end{eqnarray}
with
\begin{eqnarray}
I_{\pi N }^f &=& \frac{1}{16\pi^2}\left[-1
+\frac{p^2-m^2+M^2}{p^2}\ln\left(\frac{M}{m}\right)
+\frac{2mM}{p^2}F(\Omega)\right],
\end{eqnarray}
where
\begin{eqnarray*}
F(\Omega) &=& \left \{ \begin{array}{ll}
\sqrt{\Omega^2-1}\ln\left(-\Omega-\sqrt{\Omega^2-1}\right),&\Omega\leq -1,\\
\sqrt{1-\Omega^2}\arccos(-\Omega),&-1\leq\Omega\leq 1,\\
\sqrt{\Omega^2-1}\ln\left(\Omega+\sqrt{\Omega^2-1}\right)
-i\pi\sqrt{\Omega^2-1},&1\leq \Omega,
\end{array} \right.
\end{eqnarray*}
and
\begin{displaymath}
\Omega=\frac{p^2-m^2-M^2}{2mM}.
\end{displaymath}

\section{\label{hbchpt}HBChPT}
The expansions of the considered heavy-baryon integrals of Eq.\
(\ref{Jpnodef}) around $\Lambda =\infty$ are given by
\begin{equation}
J_{\pi N}(121;\omega)={\cal O}\left({\Lambda}\right), \label{Jpno}
\end{equation}
\begin{equation}
J_{\pi N}(021;\omega)=\frac{\Lambda^3}{16 \pi}+\frac{\Lambda^2
}{8\ \pi^2} \ \omega + {\cal O}(\Lambda),
 \label{L4Jpno}
\end{equation}
\begin{equation}
J_{\pi N}(120;\omega)=\frac{\Lambda^2}{16 \pi^2} +\ \frac{ M^2
\left[ 1+2 \ln\left( \frac{M}{\Lambda}\right)\right]}{16
\pi^2}+{\cal O}\left( \frac{1}{\Lambda^2}\right). \label{L4J012}
\end{equation}

\section{NN sector}
\subsection{Standard dimensional regularization}
   The explicit expression for the loop integral $I_{NN}$ in
dimensional regularization is given by
\begin{equation}
I_{NN}= 2\bar{\lambda} + I_{NN}^f,
\end{equation}
with
\begin{equation}
\bar{\lambda}=\frac{m^{n-4}}{16\pi^2}\left\{\frac{1}{n-4}
-\frac{1}{2}[\ln(4\pi)+\Gamma'(1)+1]\right\}
\end{equation}
and
\begin{eqnarray}
I^f_{NN} &=& -\frac{1}{16 \pi^2} \ \left[ 1+\sqrt{1-\frac{4
m^2}{P^2}} \ln\left( \frac{1-\sqrt{1-\frac{4
m^2}{P^2}}}{1+\sqrt{1-\frac{4 m^2}{P^2}}}\right)+i \pi
\sqrt{1-\frac{4 m^2}{P^2}}
\right]\nonumber\\
&=& -\frac{1}{16\ \pi^2} - \frac{i p}{16\ \pi m}+{\cal O}\left(
p^2\right).
\end{eqnarray}
The subtracted loop integral $I_{NN}^R$ reads
\begin{equation}
I_{NN}^R=I_{NN}-I_{NN}|_{P^2=4 m^2}=- \frac{i p}{16 \pi m}+{\cal
O}\left( p^2\right).
\end{equation}
For the vector integral $I_{NN}^{(P)}$ we obtain
\begin{equation}
I_{NN}^{(P)}= \bar{\lambda} + \frac{1}{2}I_{NN}^f,
\end{equation}
and the subtracted loop integral $I_{NN}^{(P) R}$ is given by
\begin{equation}
I_{NN}^{(P) R}=I_{NN}^{(P)}-I_{NN}^{(P)}|_{P^2=4 m^2}=- \frac{i
p}{32 \pi m}+{\cal O}\left( p^2\right).
\end{equation}
The tensor integral is given by
\begin{eqnarray}
I^{\mu\nu}_{NN} &=& g^{\mu\nu}\left[\frac{6m^2-P^2}{6} \
\bar{\lambda}+\frac{P^2 - 6 m^2 }{288 \pi^2}\right] + P^\mu
P^\nu\left[\frac{6 m^2-P^2}{288 \pi^2 P^2}+\frac{2}{3}\bar{\lambda}\right]\nonumber\\
&& + \left[ \frac{g^{\mu\nu} \left(4 m^2 - P^2\right)}{12} +
\frac{P^\mu P^\nu}{3} \left( 1 - \frac{m^2}{P^2}\right)\right] \
I_{NN}^f,
\end{eqnarray}
which, after subtraction, reads
\begin{equation}
I^{\mu\nu R}_{NN}=I^{\mu\nu}_{NN}-I^{\mu\nu}_{NN}|_{P^2=4 m^2}= -
\frac{i p}{64\ \pi m} \ P^{\mu}P^{\nu}+{\cal O}\left( p^2\right).
\end{equation}

\subsection{Higher-derivative formulation}
In the following, the explicit expressions for the loop integrals
in higher-derivative formulation up to and including order ${\cal
O}\left(\frac{1}{\Lambda^2}\right)$ are given.
   The scalar integral reads
\begin{equation}
I_{NN}=I_{NN}^f + \frac{1}{16 \pi^2} + \frac{1}{8 \pi^2} \
\ln\left(\frac{m}{\Lambda}\right) + \frac{1}{\Lambda^2} \left[ -
\frac{m^2}{16 \pi^2} + \frac{5 P^2}{96 \pi^2} + \frac{m^2}{4
\pi^2} \ \ln\left(\frac{m}{\Lambda}\right) \right]+{\cal
O}\left(\frac{1}{\Lambda^4}\right), \label{intINNnew}
\end{equation}
so that
\begin{equation}
I_{NN}^R=I_{NN}-I_{NN}|_{P^2=4 m^2}=- \frac{i p}{16 \pi m}+{\cal
O}\left( p^2\right). \label{NNIntexp}
\end{equation}
The vector integral is given by
\begin{equation}
I_{NN}^{(P)} = \frac{1}{2} \ I_{NN}^f + \frac{1}{32 \pi^2} +
\frac{1}{16 \pi^2} \ \ln\left(\frac{m}{\Lambda}\right) +
\frac{1}{\Lambda^2} \left[ -\frac{m^2}{32 \pi^2} + \frac{5
P^2}{192 \pi^2} + \frac{m^2}{8 \pi^2} \
\ln\left(\frac{m}{\Lambda}\right)\right], \label{intINNvec}
\end{equation}
and, after subtraction, one obtains
\begin{equation}
I_{NN}^{(P) R} =I_{NN}^{(P)}-I_{NN}^{(P)}|_{P^2=4 m^2}= -\frac{i
p}{32 \pi m}+{\cal O}\left( p^2\right). \label{intINNexp}
\end{equation}
The expression for the tensor integral reads
\begin{eqnarray}
I^{\mu\nu}_{NN} &=& \frac{g^{\mu\nu}}{64 \pi^2} \ \Lambda^2 +
\frac{g^{\mu\nu}\ \left[ m^2 + \left( P^2-6 m^2
\right)\ln\left(\frac{\Lambda}{m}\right)\right]}{96 \pi^2}
+\frac{P^\mu P^\nu \left[ 1 +\frac{m^2}{p^2}
-2 \ln\left(\frac{\Lambda}{m}\right)\right]}{48 \pi^2}\nonumber\\
&& +\left[ \frac{g^{\mu\nu}\ \left(4 m^2 - P^2\right)}{12} +
\frac{P^\mu
P^\nu}{3} \left( 1 - \frac{m^2}{P^2}\right)\right] \ I_{NN}^f \nonumber\\
&& + \frac{1}{\pi^2 \Lambda^2} \ \Biggl\{  \left[ -\frac{m^4}{32}
+ \frac{5 m^2 P^2}{192} -\frac{(P^2)^2}{480}-\frac{m^4}{16} \
\ln\left(\frac{\Lambda}{m}\right)\right] g^{\mu\nu}\nonumber\\
&& + P^\mu P^\nu \left[ \frac{m^2}{48} + \frac{17 P^2}{960}-
\frac{m^2}{8} \ \ln\left(\frac{\Lambda}{m}\right)\right] \Biggr\},
\end{eqnarray}
and
\begin{equation}
I^{\mu\nu R}_{NN}=I^{\mu\nu}_{NN}-I^{\mu\nu}_{NN}|_{P^2=4 m^2}= -
\frac{i p}{64\ \pi m} \ P^{\mu}P^{\nu}+{\cal O}\left( p^2\right).
\end{equation}

\end{document}